\documentclass[10pt,a4paper,twocolumn]{article}
\usepackage{euler,amsmath,defns}
\IfFileExists{ajr.sty}{\usepackage{ajr}}{\usepackage{graphicx,url,hyperref}}
\allowdisplaybreaks

\newcommand{\uu}{{\bar u}}

\newcommand{\ros}{{\dot\varepsilon}}
\newcommand{\vel}{{\vec u}}

\title{Model turbulent floods with the Smagorinski large eddy closure}

\author{A.~J. Roberts\thanks{Computational Engineering and Science
Research Centre, Department of Mathematics \& Computing, University of
Southern Queensland, Toowoomba, Queensland 4352, Australia.
\protect\url{mailto:aroberts@usq.edu.au}}
\and D.~J. Georgiev\footnotemark[1] 
\and D.~V. Strunin\footnotemark[1]}

\begin{document}
    
\maketitle
    
\begin{abstract}
Floods, tides and tsunamis are turbulent, yet conventional models are based upon depth averaging inviscid irrotational flow equations.  We propose to change the base of such modelling to the Smagorinksi large eddy closure for turbulence in order to appropriately match the underlying fluid dynamics.   Our approach allows for large changes in fluid depth to cater for extreme inundations.  The key to the analysis underlying the approach is to choose surface and bed boundary conditions that accommodate a constant turbulent shear as a nearly neutral mode.    Analysis supported by slow manifold theory then constructs a model for the coupled dynamics of the fluid depth and the mean turbulent lateral velocity.  The model resolves the internal turbulent shear in the flow and thus may be used in further work to rationally predict erosion and transport in turbulent floods.
\end{abstract}

\paragraph{Keywords:} turbulent flood, tsunami, Smagorinski closure, shallow water equations

\paragraph{PACS:} 47.27.E-, 47.27.nd, 47.10.Fg, 47.11.St, 92.10.Sx


\section{Introduction}

This letter models the two~dimensional turbulent flow of a layer of fluid over flat sloping ground.  The fluid of thickness~$\eta(x,t)$ flows with mean lateral velocity~$\uu(x,t)$.  Following earlier but more complex modelling using the $k$-$\epsilon$ model of turbulence~\cite{Roberts99c}, we approximate the turbulence within the fluid by the nonlinear constitutive relation of the Smagorinski closure \cite[e.g.]{Kim02, Ozgokmen07}.  Then the systematic analysis developed in this letter recommends using a nondimensional turbulent model described by
\begin{align} 
    \D t\eta +\D x{}\left[ \eta\uu \right] = {}&0\,,
    \label{eq:ghup0}
    \\
    \D t\uu +1.05\,\uu\D x\uu \approx {}&
    -0.0031\frac{|\uu|\uu}\eta
    +0.98\left(g_x-g_z\D x\eta\right)
    \nonumber\\&{}
    +0.26\frac{|\uu|^{0.78}}\eta\D x{}
    \left( \eta^2|\uu|^{0.22}\D x\uu\right) \,,
    \label{eq:gup0}
\end{align}
where $g_x$~and~$g_z$ are the nondimensional components of gravity along and normal to the flat ground, respectively.  Fluid is conserved through~\eqref{eq:ghup0}.  The mean momentum equation~\eqref{eq:gup0} incorporates effects of inertia,~$\uu_t$, self-advection,~$\uu\uu_x$, bed drag,~$|\uu|\uu/\eta$, gravitational forcing, $(g_x-g_z\eta_x)$, and an enhanced lateral turbulent mixing described by the last term.  For example, for turbulent flow down an inclined flat plate with lateral gravity~$g_x$, the nonlinear bed friction may balance gravitational forcing whence the above model predicts the equilibrium flow to have mean lateral velocity $\uu\approx 18\sqrt{g_x\eta}$\,.   The model~\eqref{eq:ghup0}--\eqref{eq:gup0} also resolves instabilities from this equilibrium flow and the emergence and interaction of solitary or roll waves on the turbulent flow.  In a personal communication, Howell Peregrine commented that environmental shallow water waves were observed to propagate 1--2\% slower than predicted by conventional shallow water theory; the above turbulent model successfully predicts just such slight slowing because of the slightly reduced effect of gravity in the coefficent~$0.98$ in~\eqref{eq:gup0}.   

This letter puts the model \eqref{eq:ghup0}--\eqref{eq:gup0} within the sound support of modern dynamical systems theory, Section~\ref{sec:smwcmt}, to empower us to systematically control error, assess domains of validity, comprehensively account for further physical effects, and resolve internal structures within the turbulent flow.  
For example, Harris et al.~\cite{Harris01} modelled particle driven, gravity currents using shallow water equations that resolve the dynamics of both the fluid thickness and the mean lateral velocity.  However, such modelling of turbulent dissipative flows from the laminar inviscid foundation of shallow water equations appears a contradiction that demands resolution using the sort of approach introduced here.  
Future developments using this approach may straightforwardly incorporate complex physical effects such as the three dimensional flow over varying terrain similar to previous modelling of laminar viscous flow~\cite{Zhenquan98}.

\section{Differential equations of the Smagorinski large eddy closure}
\label{sec:demf}

Let the incompressible fluid have thickness~$\eta(x,t)$ above the ground, constant density~$\rho$,  and  flow with some turbulent mean velocity field~$\vel =(u,w)=(u_1,u_3)$ and turbulent mean pressure field~$p$. In this letter we restrict attention to two dimensional turbulent flow.

\paragraph{Smagorinski large eddy closure} Define the turbulent mean strain-rate tensor \cite[e.g.]{Stickel05}
\begin{equation}
	\ros_{ij} :=\rat12\left( \D{x_j}{u_i} +\D{x_i}{u_j} \right) \,,
\end{equation}
where $x_1=x$ and $x_3=z$ are distances along and normal to the mean solid substrate, respectively.  Then take the turbulent mean stress tensor for the fluid to be $\sigma_{ij} =-p\delta_{ij} +2\rho\nu \ros_{ij}$ for some effective turbulent eddy viscosity~$\nu$. The Smagorinksi model~\cite[e.g.]{Ozgokmen07} for very large Reynolds number flows then sets the turbulent eddy viscosity~$\nu$ to be linear in the magnitude~$\ros$ of the second invariant of the strain-rate tensor:
\begin{equation} 
\nu:=c_t\eta^2\ros
\qtq{where}
    \ros^2:=\sum_{i,j}\ros_{ij}^2\,.
    \label{eq:cons}
\end{equation}
Comparison with established channel flow experiments~\cite[e.g.]{Nezu05}, Section~\ref{sec:cfcm}, indicates the constant of proportionality $c_t\approx  0.02$ for the turbulent environmental flows we consider.

The lateral turbulent mixing in the model~\eqref{eq:gup0}, strength $0.26\eta\uu$, is an order of magnitude larger than direct mixing in the underlying Smagorinski model, $\nu_t\approx 0.02\eta^2\D zu$\,, because of the `shear dispersion' enhancement to lateral mixing via the interaction between lateral shear and vertical mixing.  Depth averaging misses this important physical interaction.  This enhancement in~\eqref{eq:gup0} qualitatively agrees with Bijvelds et al.~\cite{Bijvelds99} who increased the lateral eddy viscosity in order for their numerical model to match laboratory experiments.

\paragraph{Partial differential equations} 

Make variables nondimensional with respect to some velocity scale, a typical fluid thickness, and the fluid density.  Then the nondimensional
\pde{}s for the incompressible, two~dimensional, turbulent mean, fluid flow are firstly the continuity equation
\begin{equation}
    \divv\vel =\D xu+\D yv=0\,,
\end{equation}
and secondly the momentum equation
\begin{equation}
     \D t{\vel } +\vel \cdot\grad\vel 
    =-\grad p +\divv\vec\tau +\vec{ g}\,,
    \label{eq:mom}
\end{equation}
where  $\vec g=(g_x,g_z)$ is the nondimensional forcing of gravity, and $\vec\tau$~is the nondimensional, turbulent mean, deviatoric, stress tensor:\footnote{Some~\cite[e.g.]{Kim02} like to use the negative of this stress tensor, but then ``$+\divv\vec\tau$'' in~\eqref{eq:mom} has to be ``$-\divv\vec\tau$''.}
\begin{equation}
    \tau_{ij}=2\nu(\ros)\ros_{ij}
    =c_t\eta^2\ros \left( \D{x_j}{u_i} +\D{x_i}{u_j} \right)\,.
\end{equation}

\paragraph{Boundary conditions} Compatible with the Smagorinski closure, we formulate boundary conditions on the ground in terms of the turbulent mean velocity and the fluid depth. On the mean bed allow slip to account for a relatively thin turbulent boundary layer,
    \begin{equation}
    w=0 \qtq{and} u=c_u\eta \D zu \quad\text{on }z=0\,,
        \label{eq:noslip}
    \end{equation}
    where comparison with experiments on channel flow, Section~\ref{sec:cfcm}, indicates the coefficient $c_u\approx 1.85$\,; in applications this coefficient would vary to parametrise different ground roughness. We envisage  that $z=0$ denotes the local \emph{mean} level of the ground in analogy to $\vel$~and~$p$ denoting the \emph{turbulent mean} velocity and pressure fields.

The kinematic condition on the turbulent mean free surface is 
    \begin{equation}
        \D t\eta+u\D x\eta=v\quad\text{on } z=\eta\,,
    \end{equation}
    where similarly $\eta$~denotes the \emph{turbulent mean} location of the free surface.

The turbulent mean stress normal to the free surface comes from constant environmental pressure, taken to be zero, that is,
    \begin{equation}
		-p+\frac1{1+\eta_x^2}\left(\tau_{33} -2\eta_x\tau_{13}
		+\eta_x^2\tau_{11} \right)
		= 0
		 \quad\text{on } z=\eta\,.
    \end{equation}
    
Lastly, there must be no turbulent mean, tangential, stress at the free surface:
    \begin{equation}
		(1-\eta_x^2)\tau_{13}+\eta_x(\tau_{33}-\tau_{11})=0
		\quad\text{on } z=\eta\,.
        \label{bc:tt}
    \end{equation}

\section{Centre manifold theory supports the modelling}
\label{sec:smwcmt}

This section describes the new approach to placing models such as
\eqref{eq:ghup0}--\eqref{eq:gup0} on a sound theoretical base.  We aim the mathematics to represent the dominant physics: as indicated by Figures 14--15 by Janosi~\cite{Janosi04}, turbulence does mix across the fluid layer. Thus we expect to be able to model the dynamics in terms of depth averaged quantities.  But instead of crudely depth averaging the equations, we use centre manifold theory to resolve turbulent dynamics across the fluid layer and hence provide a sound macroscale closure for the relatively slow, long term, dynamics of interest to environmental modellers.

To apply the theory we artificially modify the zero tangential stress free surface condition~\eqref{bc:tt} to have an artificial forcing proportional to the local velocity, a forcing parametrised by~$\gamma$ and which we later remove by evaluating at parameter $\gamma=1$ in order to recover the physical turbulent equations: on $z=\eta$ replace~\eqref{bc:tt} with
\begin{equation}
    \left[
    (1-\eta_x^2)\tau_{13}+\eta_x(\tau_{33}-\tau_{11})
    \right] 
    = \frac{(1-\gamma)c_t}{\sqrt2(1+c_u)^2} u^2 \,.
    \label{bc:tta}
\end{equation}
Evaluated at $\gamma=1$ this artificial right-hand side becomes zero so the boundary condition~\eqref{bc:tta} reduces to the physical boundary condition of zero, turbulent mean, tangential stress~\eqref{bc:tt}.  However, when the parameter $\gamma=0$ and the lateral gravity and lateral derivatives are negligible, $ g_x=\partial_x=0$\,, a neutral mode of the dynamics is the lateral shear flow $u\propto c_u+z/\eta$\,.   This neutral lateral shear mode arises because in pure shear flow the stress component $\tau_{13}=\nu \partial u/\partial z$ and hence the artificial free surface condition~\eqref{bc:tta} reduces to $\nu \partial u/\partial z =\nu u/\eta$ on $z=\eta$\,.  

Conservation of fluid provides a second neutral mode in the dynamics.  That is, when $\gamma= g_x=\partial_x=0$\,, then a two parameter family of equilibria exists corresponding to some uniform lateral shear, turbulent mean, flow, $u\propto c_u+z/\eta$\,, on a fluid of any constant thickness~$\eta$.  For large enough lateral length scales, these equilibria occur independently at each location~$x$~\cite[e.g.]{Roberts88a, Roberts96a} and hence the space of equilibria are in effect parametrised by $\uu(x)$~and~$\eta(x)$. Provided we can treat lateral derivatives~$\partial_x$ as a modifying influence, that is provided solutions vary slowly enough in~$x$, centre manifold theorems~\cite[e.g.]{Carr81, Kuznetsov95} assure us of three vitally important properties: 
\begin{enumerate}
\item this subspace of equilibria is perturbed to a slow manifold, on which the evolution is slow, that \emph{exists for a finite range} of gradients~$\partial_x$, and parameters $\gamma$~and~$ g_x$, and which may be parametrised by the average, turbulent mean, lateral velcity~$\uu(x,t)$ and the local thickness of the fluid~$\eta(x,t)$; 

\item the slow manifold \emph{attracts solutions from all nearby initial conditions} exponentially quickly on a cross-depth mixing time; and that 

\item \label{en:3} a formal power series in the parameters~$\gamma$, $ g_x$ and gradients~$\partial_x$ \emph{approximates} the slow manifold to the same order of error as the order of the residuals of the governing differential equations.\end{enumerate}
That is, the theorems support the existence, relevance and construction of slow manifold models such as \eqref{eq:ghup0}--\eqref{eq:gup0}.

An alternative and powerful view of these theorems is that they follow from a nonlinear, normal form, coordinate transform that decouples the slow and fast modes in the fluid dynamics~\cite[e.g.]{Roberts04b}.  That is, the models we discuss are essentially simply a reparametrisation of the state space, but only that part of the state space where the relatively slow dynamics occurs.

\section{Low order models of the dynamics}
\label{sec:lomd}

The detailed and complicated algebra deriving the model is of little interest to users of the model. Computer algebra readily constructs our slow manifold models~\cite{Roberts08e}: those interested could check the code and verify that the algorithm asymptotically solves the governing differential equations and boundary conditions as specified~\cite[pp.9--16]{Roberts08e}.  The asymptotic construction of the slow manifold is valid for small lateral derivatives, small lateral forcing and small perturbation of the free surface condition [Property~\ref{en:3}]. This section focusses on the resulting model and its interpretation.

\subsection{Channel flow calibrates the model}
\label{sec:cfcm}

Open channel flow is well studied.
Experiments give empirical structure of the turbulence in the interior of the flow which we use to calibrate the two constants in our Smagorinski model of turbulent floods.

The cross fluid structure of the velocity, pressure and strain rate forms the slow manifold, here with no lateral variations, $\partial_x=0$\,, as we compare with uniform flow.  In terms of the scaled normal coordinate $\zeta=z/\eta$\,, computer algebra~\cite{Roberts08e} deduces the lateral shear velocity profile 
\begin{align}
u= {}&
\uu {\textstyle \frac{2(\zeta+c_u)}{1+2c_u} }
\nonumber\\&{}
+\gamma\uu \textstyle 
\frac{(1+c_u)[ (1+4c_u)(c_u+\zeta) -2(1+2c_u)(3c_u\zeta^2 +\zeta^3) ]} {4(1+2c_u)^2(1+3c_u+3c_u^2)}
\nonumber\\&{}
+\frac{g_x\eta}{\uu}\,  \textstyle 
\frac{[(5+6c_u)(c_u+\zeta) -6(2+7c_u+6c_u^2)\zeta^2 +6(1+2c_u)^2\zeta^3]} {48\sqrt2 \, c_t(1+3c_u+3c_u^2)}
\nonumber\\&{}
+\Ord{\gamma^2+g_x^2+\partial_x} \,,
\label{eq:upro}
\end{align}
the magnitude of the rate of strain tensor is
\begin{align}
\ros = {}& 
\frac{\uu}{\eta}  {\textstyle \frac{\sqrt2}{1+2c_u} }
+\frac{\gamma\uu}{\eta}\, \textstyle 
\frac{\sqrt2(1+c_u)[(1+4c_u)-6(1+2c_u)(2c_u\zeta+\zeta^2)]}
{8(1+2c_u)^2(1+3c_u+3c_u^2)}
\nonumber\\&{}
+\frac{g_x}{\uu}\, \textstyle 
\frac{[(5+6c_u)-12(2+7c_u)\zeta +18(1+2c_u)^2\zeta^2]}{96c_t(1+3c_u+3c_u^2)}
\nonumber\\&{}
+\Ord{\gamma^2+g_x^2+\partial_x} \,,
\label{eq:rosux}
\end{align}
and flows with hydrostatic pressure, $p=g_z\eta(1-\zeta)$\,.
These fields approximate out-of-equilibrium turbulent flow.  The corresponding evolution of the mean lateral velocity is
\begin{align}
\frac{d\uu}{dt}= {}& { \textstyle 
-\frac{\sqrt2\,3c_t(1+c_u)}{(1+2c_u)(1+3c_u+3c_u^2)} }
\frac{\gamma \uu^2}\eta
\textstyle 
+\frac{\rat34+3c_u+3c_u^2}{1+3c_u+3c_u^2}g_x
\nonumber\\&{}
+\Ord{\gamma^2+g_x^2+\partial_x} 
\label{eq:gux}
\end{align}
Obtain a physical model by setting the artificial parameter $\gamma=1$\,.
Lateral gravitational forcing, the second term, accelerates the flow until balanced by dissipation, the first term, to reach the well known equilibrium for open channel flow.

\begin{figure}
\centering
\begin{tabular}{c@{}c}
\rotatebox{90}{\hspace{15ex}$\zeta=z/\eta$}&
\includegraphics[width=0.9\linewidth]{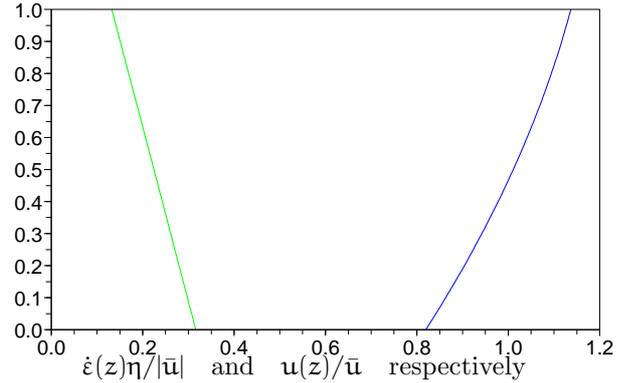}\\[-2ex]
& $\ros(z) \eta/|\uu| \qtq{and} u(z)/\uu$\quad respectively
\end{tabular}
\caption{approximate vertical profiles at open channel flow equilibrium for which $g_x\eta/\uu^2=0.0031$\,.}
\label{fig:vp}
\end{figure}

Observations calibrate the Smagorinski coefficient~$c_t$ and bed `slip' coefficient~$c_u$.
Many experiments confirm the equilibrium flow velocity is $\uu=18u_*$ for  `friction velocity' $u_*:=\sqrt{g_x\eta}$\,.
Summarising many experiments,  Nezu gave empirical formulae for the structure of turbulent energy~$k$ and dissipation~$\epsilon$ \cite[(17) and (21)]{Nezu05}; we integrate the eddy viscosity $\nu(z)=C_\mu k^2/\epsilon$ obtained from these formulae over the depth gives the empirical, depth averaged, eddy viscosity $\bar\nu=0.0796\,\eta u_*$\,.  To match the equilibrium flow velocity and the average eddy viscosity we find $c_t=1/50$ and $c_u=11/6$  which we use hereafter.\footnote{The equations determining these constants appear to be a reasonably well conditioned.}  Figure~\ref{fig:vp} plots the corresponding vertical profiles of the rate-of-strain~$\ros$ and lateral shear velocity~$u$ at the equilibrium of open channel flow: the lateral velocity shows a small shear corresponding to the turbulent mixing across the fluid with dissipation at the bed; the rate-of-strain increases towards the bed (until the bottom boundary layer which is not resolved in Figure~\ref{fig:vp}).  Our resolution of these internal fields could contribute to prediction of erosion and deposition in further modelling.

\subsection{Turbulent spatio-temporal dynamics} 

The previous subsection briefly explored the equilibrium of uniform open channel flow.  Now we address modelling flood flow with mean turbulent properties varying in space and time.

Further computer algebra~\cite{Roberts08e} generates modifications to cross fluid structures such as \eqref{eq:upro}--\eqref{eq:gux} to account for relatively slow variations in the lateral direction~$x$, small but non-zero~$\partial_x$.  This letter is no place to record the considerable algebraic detail.  However, the result is that a model for the evolution of mean lateral velocity~$\uu(x,t)$ is
\begin{align}
\D t\uu =  {}&
-1.045\,\uu\D x\uu
-0.00311\,\gamma\frac{|\uu|\uu}\eta
+0.985\left(g_x-g_z\D x\eta\right)
\nonumber\\&{}
+0.058\,\eta\left(\D x\uu\right)^2
+0.259\,\eta|\uu|\DD x\uu
+0.522\,|\uu|\D x\eta\D x\uu
\nonumber\\&{}
-0.015\frac{\eta_x}\eta\uu^2
-0.007\frac{\eta_x^2}\eta\uu^2
-0.007\,\uu^2\DD x\eta
\nonumber\\&{}
+\Ord{\gamma^{3/2}+g_x^{3/2}+g_z^3+\partial_x^3}\,.
\label{eq:guf}
\end{align}
This would be solved with the associated equation~\eqref{eq:ghup0} representing conservation of fluid.  
Obtain the model \eqref{eq:ghup0}--\eqref{eq:gup0} discussed in the Introduction of this article by four steps:
evaluate~\eqref{eq:guf} at the physically relevant $\gamma=1$ to remove the artifice in the surface boundary condition~\eqref{bc:tta}; approximate coefficients to two decimal places; neglect the numerically small effects recorded in the third line of~\eqref{eq:guf}; and, to small numerical errors, combine the three terms on the second line of~\eqref{eq:guf} into the one nonlinear lateral mixing term on the second line of~\eqref{eq:gup0}.  Importantly, although the models \eqref{eq:gup0}~and~\eqref{eq:guf} are expressed in terms of depth averaged lateral velocity, remember that they are derived, not by depth averaging, but instead by systematically accounting for interactions between vertical profiles of velocity\slash stress and bed drag and lateral space variations.  Centre manifold theory provides the framework to account for these physical processes and so deduce these physical models.  In particular, this approach provides a sound lateral regularising dissipation in the second lines of \eqref{eq:gup0}~and~\eqref{eq:guf}.

Computer algebra~\cite[\S5]{Roberts08e} constructs terms in the formal power series solutions in the notionally small parameters~$\gamma$, $g_x$, $g_z$ and~$\partial_x$.   We record \emph{one} truncation of the formal power series in~\eqref{eq:guf}.  Other truncations of the multivariate power series generate other valid approximations of varying accuracy.    With the support of centre manifold theory, researchers may choose an approximate model that suits the parameter regime of their application.

\subsection{Turbulent roll waves arise}

\begin{figure}
\centering
\begin{tabular}{c@{}c}
\rotatebox{90}{\hspace{5ex}$\eta-1\qtq{and}\uu-3.55$}&
\includegraphics[width=0.9\linewidth]{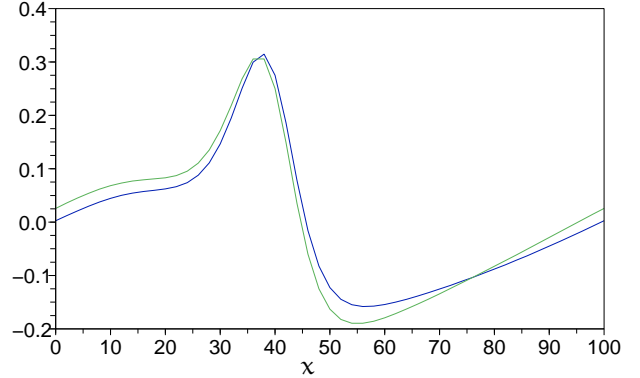}\\[-2ex]
&$x$
\end{tabular}
\caption{turbulent roll wave on a 4\%~slope on a fluid of nondimensional mean depth~$1$ and with nondimensional mean lateral flow velocity~$3.55$.  The fluid depth (blue) varies by nearly~50\% and is closely in phase with the velocity variations (green).}
\label{fig:smagsimhu}
\end{figure}

Roll waves spontaneously arising on fluid flowing down a slope have been noted and analysed for a century~\cite[e.g.]{Needham84, Balmforth04}.  Figure~\ref{fig:smagsimhu} plots the profile of one period of such a \emph{turbulent} roll wave on a slope as predicted by the model \eqref{eq:ghup0}--\eqref{eq:gup0}.  The occurrence of such roll waves is of concern for hydraulic engineers: the model \eqref{eq:ghup0}--\eqref{eq:gup0} predicts turbulent roll waves arise on slopes greater than~1.38\%.

\begin{figure}
\centering
\begin{tabular}{cc}
\rotatebox{90}{\hspace{15ex}slope $g_x$}&
\includegraphics[width=0.9\linewidth]{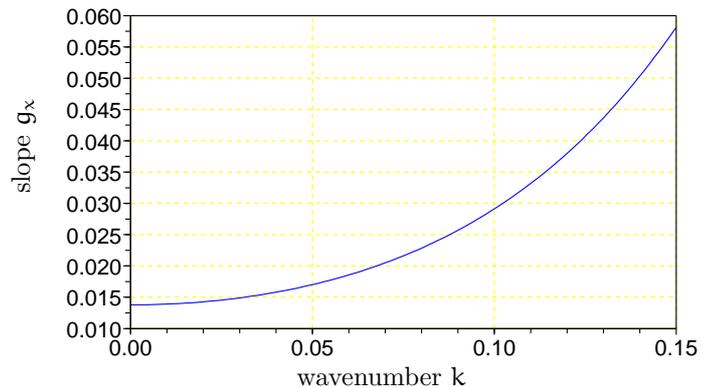}\\[-1ex]
&wavenumber $k$
\end{tabular}
\caption{critical slope~$g_x$ for the appearance of turbulent roll waves of wavenumber~$k$ as predicted by model \eqref{eq:ghup0}--\eqref{eq:gup0}.}
\label{fig:rollwaves}
\end{figure}

Standard stability analysis on a fluid of nondimensional depth~$1$ and the corresponding nondimensional equilibrium flow velocity~$17.8\sqrt{g_x}$ seeks small perturbations $\epsilon\exp(ikx+\lambda t)$\,.  For model \eqref{eq:ghup0}--\eqref{eq:gup0} the characteristic equation for such perturbations is 
\begin{align*}&
\lambda^2
-(0.1102+36.5ik+4.623k^2)\sqrt{g_x}\lambda
\\&{}
+\big[2.94g_xik+(0.98-331.9g_x)k^2+82.19g_xik^3\big]
=0\,.
\end{align*}
Consequently, the real part of the growth rate~$\lambda$ crosses zero for critical slopes 
\begin{equation}
g_x^c=\frac{0.00296+0.248k^2+5.2k^4}{0.2146-k^2}\,,
\end{equation}
as plotted in Figure~\ref{fig:rollwaves} (note that the nondimensional~$g_x$ effectively denotes the slope of the bed).  Roll wave instability arises first for long waves on slopes steeper than~1.38\%.  On a slope of~4\%, as simulated in Figure~\ref{fig:smagsimhu}, roll waves are unstable for nondimensional wavenumber $|k|<0.12$\,, that is, for wavelengths longer than~$52\eta$.  These predictions are experimentally verifiable.

\section{Conclusion}

Following similar modelling for viscous thin fluid films~\cite{Roberts99b, Roberts07b}, the innovation of modifying the free surface condition to~\eqref{bc:tta} with the Smagorinski closure places the modelling of turbulent fluid flows upon the powerful and sound basis of centre manifold theory~\cite[e.g.]{Carr81, Kuznetsov95}.  Such a modern dynamical system foundation empowers us to systematically derive the models~\eqref{eq:ghup0}, \eqref{eq:gup0} and~\eqref{eq:guf} for turbulent tides, floods and tsunamis.  Earlier modelling based upon the $k$-$\epsilon$ turbulence model led to a description of floods in terms of the depth averaged velocity, turbulent energy and turbulent dissipation~\cite{Roberts99c}.  The extra complexity of these extra field variables makes the earlier model more problematical to use in forecasting environmental flows.  The simpler model here should prove more useful.

The derivation resolves vertical profiles within the flow, and the physical processes and interactions between these profiles and lateral variations.  Thus these models, via \eqref{eq:upro}~and~\eqref{eq:rosux} for example, can be used to rationally predict bed strain and hence erosion and transport in detail for out of equilibrium flows. Further refinement of the computer algebra derivation~\cite{Roberts08e} will empower sound modelling of three dimensional turbulent flows,  via the Smagorinski closure, on varying topography as already done for viscous fluid films~\cite{Roberts99b}.

\bibliographystyle{plain}
\bibliography{ajr,bib}

\end{document}